\documentclass[prd,twocolumn,amsmath,nobibnotes,nofootinbib]{revtex4}

\usepackage{bm} \usepackage{graphicx} \usepackage{epstopdf}
 \def\ba{\begin{eqnarray}} \def\ea{\end{eqnarray}}
\def\be{\begin{equation}} \def\ee{\end{equation}} \def\({\left(}
\def\){\right)} \def\[{\left[} \def\]{\right]} \def\<{\left<}
\def\>{\right>}

\def\ba{\begin{eqnarray}}
\def\ea{\end{eqnarray}}
\def\be{\begin{equation}}
\def\ee{\end{equation}}
\def\({\left(}
\def\){\right)}
\def\[{\left[}
\def\]{\right]}
\def\<{\left<}
\def\>{\right>}

\begin{document}

\title{Towards observable signatures of other bubble universes}
\date{\today}

\author{Anthony Aguirre}
\email{aguirre@scipp.ucsc.edu}
\affiliation{SCIPP, University of California, Santa Cruz, CA 95064, USA}
\author{Matthew C Johnson}
\email{mjohnson@physics.ucsc.edu}
\affiliation{SCIPP, University of California, Santa Cruz, CA 95064, USA}
\author{Assaf Shomer}
\email{shomer@scipp.ucsc.edu}
\affiliation{SCIPP, University of California, Santa Cruz, CA 95064, USA}

\begin{abstract}
We evaluate the possibility of observable effects arising from collisions between vacuum bubbles in a universe undergoing false-vacuum eternal inflation. Contrary to conventional wisdom, we find that under certain assumptions most positions inside a bubble should have access to a large number of collision events. We calculate the expected number and angular size distribution of such collisions on an observer's ``sky," finding that for typical observers the distribution is anisotropic and includes many bubbles, each of which will affect the majority of the observer's sky. After a qualitative discussion of the physics involved in collisions between arbitrary bubbles, we evaluate the implications of our results, and outline possible detectable effects. In an optimistic sense, then, the present paper constitutes a first step in an assessment of the possible effects of other bubble universes on the cosmic microwave background and other observables.
\end{abstract}

\maketitle

\section{Introduction}
\label{sec-intro}

Cosmological inflation never ends globally when driven by an inflaton potential with  long-lived metastable minima.  This was discovered in the very first models of inflation as a failure of ``true" (lower) vacuum bubbles in a ``false" vacuum background to percolate~\cite{Guth:1982pn}. It was later recognized as a special case of ``eternal inflation" in which our observable universe would lie within a single nucleated bubble~\cite{Gott:1982zf} while inflation continues forever outside of this bubble (e.g.,~\cite{Guth:1982pn,Vilenkin:1992uf}).

While important for any sufficiently complicated inflaton potential, this issue has become prominent lately with the realization that stabilized string theory compactifications appear to correspond to minima of a many-dimensional effective potential ``landscape"~\cite{Susskind:2003kw,Douglas:2003um} that would drive just this sort of eternal inflation and thus create ``pocket" or ``bubble" universes with diverse properties. This has raised a number of very thorny questions regarding which properties to compare to our local observations (e.g.~\cite{Aguirre:2004qb,Hartle:2004qv}), as well as debates as to whether these other ``universes" have any meaning if they are unobservable, as is the conventional wisdom.

But what if they {\em are} observable, so that the processes responsible for eternal inflation can be directly probed? What is the chance we could actually see such bubbles, and how would they look on the sky?  These are the questions that the present paper begins to explore.

It would seem that for us to observe bubble collisions in our past, three basic and successive criteria must be met:

\begin{enumerate}

\item{{\em Compatibility:} A bubble collision must allow standard cosmological evolution including inflation and reheating -- and hence be potentially compatible with known observations -- in at least part of its future lightcone.}

\item{{\em Probability:} Within a given ``observation bubble"  (seen as a negatively-curved Friedmann-Robertson-Walker (FRW) model by its denizens) a randomly chosen point in space should have a significant probability of having (compatible) bubbles to its past.}

\item{{\em Observability:} The effects of compatible bubbles to the past must not be diluted away by inflation into unobservability, nor affect a negligible area of the observer's sky.}

\end{enumerate}

Although a rigorous analysis of these issues does not yet exist, several recent studies suggest -- in contrast to previous thinking -- that it is actually plausible that these three criteria may be met.

First, studies of bubble collisions ``boosted" so that one bubble forms much ``earlier" than the other indicate that the older bubble may see the younger bubble as a small perturbation that does not disrupt its overall structure~\cite{Bousso:2006ge}, even if the younger bubble contains a big-crunch singularity~\cite{Freivogel:2007fx}.
Second, straightforward arguments (see below), inspired by the results of Garriga, Guth \& Vilenkin~\cite{Garriga:2006hw} (hereafter GGV), indicate that a random position in the FRW space within a bubble should (with probability one) have a bubble nucleation event to its past. Third, in a complex inflaton potential with many minima, the number of e-foldings within a randomly chosen bubble can become a random variable with some probability distribution.  Suppose that this distribution favors a small number of e-foldings, and yet -- either to match our observations or for ``anthropic" reasons -- we focus only on the subset of bubbles with $\agt N_{\rm min}\sim 50-60\,$e-foldings.  Then we might expect that our region underwent close to $N_{\rm min}$ e-foldings~\cite{Tegmark:2004qd,Freivogel:2005vv}. Thus it is plausible that just enough inflationary e-foldings occurred to explain the largeness and approximate flatness of the universe; and since the CMB perturbations on the largest scales formed $\sim N_{\rm min}$ e-foldings before the end of inflation, perturbations at the beginning of inflation may then be detectable.

None of these studies have actually addressed whether bubble
collisions might be observable, however, and leave many key
questions unresolved. The bulk of the present paper aims to help
answer several of these questions by calculating, given an observer
at an arbitrary spacetime point in a bubble, the expected
differential number
\begin{equation}\label{difn}
\frac{dN}{d\psi d(\cos\theta) d\phi}
\end{equation}
of bubble collisions on the observer's bubble wall, seen on the sky by the observer with angular scale $\psi$ and direction $(\theta,\phi)$.

We will see that for small nucleation rates, this distribution is
interesting for two cases.  First, very late-time observers might
see a nearly-isotropic distribution of bubbles with tiny angular
scales. Second, for a typical position inside the bubble, many
bubbles enter the past lightcone at early times and with large
angular scales (i.e., each collision will affect the majority of the
observer's sky), nearly all from a particular direction on the sky.
While we can only speculate as to how these bubbles would look
observationally, the detection of either signal would offer direct
observational evidence that we inhabit a universe undergoing
false-vacuum eternal inflation, and would bolster support for
fundamental theories that may drive this type of cosmological
evolution.

We proceed as follows. In Sec.~\ref{sec-setup} we discuss the dS background and the structure of a bubble universe inside it, then outline the calculation to be performed and the simplifying assumptions we will employ.  In Sec.~\ref{sec-computations} we display the calculation. The basic results and their implications are summarized in Sec.~\ref{sec-resimp}, and readers uninterested in the details of the computations can skip from Sec.~\ref{sec-setup} to this section. Finally, in Sec.~\ref{sec-discuss} we conclude.

\section{Setting up the problem}
\label{sec-setup}

The system we will study consists of a de Sitter spacetime (dS) supported by a false-vacuum energy, containing nucleated Coleman-de Luccia (CDL)~\cite{Coleman:1980aw, Coleman:1977py,Callan:1977pt} bubbles of true vacuum. We work in the approximation where all bubbles are nucleated with vanishing size, expand at the speed of light, and have an infinitely thin wall. Bubble walls  then correspond to spherically symmetric null shells.

The geometry of the bubble interior, the background de Sitter space, and the wall between them can all be visualized and understood in terms of a 5D embedding space with coordinates $X^{\mu}$, $\mu=0...4$, and Minkowski metric $ds^2=\eta_{\mu\nu}dX^{\mu} dX^{\nu}$. In this embedding, pure dS is a hyperboloid defined by $\eta_{\mu\nu}X^{\mu} X^{\nu} = H^{-2}$, where $H^2=8\pi\rho_\Lambda/3$ in terms of the vacuum energy density $\rho_\Lambda$.

In formulating the problem we employ the ``flat slicing" coordinates $(t,r,\theta,\phi)$ to describe the dS (with $H=H_F$) outside of the bubble. In the embedding space, these coordinates are given by
\begin{eqnarray}
&& X_{0} = H_{F}^{-1} \sinh H_F t + \frac{H_F}{2}e^{H_F t}r^2 \\ \nonumber
&& X_{i} = r e^{H_F t} \omega_i \\ \nonumber
&& X_{4} = H_{F}^{-1} \cosh H_F t - \frac{H_F}{2}e^{H_F t}r^2,
\end{eqnarray}
with $(\omega_1,\omega_2,\omega_3)=(\cos\theta,\sin\theta\cos\phi,\sin\theta\sin\phi)$, $0 \le r < \infty$, $-\infty < t < \infty$.  This induces the metric:
\begin{equation}\label{flatds}
ds^2 = -d t^2 +e^{2H_F t} \left[d r^2 + r^2 \ d\Omega_{2}^{2} \right],
\end{equation}
which covers half of the de Sitter hyperboloid.

Turning now to the bubble, the  exact form of the post-nucleation bubble interior is found from the analytic continuation of the CDL instanton~\cite{Coleman:1980aw}, with the details largely dependent on the form of the inflaton potential. The null cone, which in our approximation traces the wall trajectory, more generally corresponds to the post-tunneling field value.\footnote{At late times, the identification of the null cone with the position of the bubble wall becomes an increasingly accurate approximation, and we can safely neglect the portion of the spacetime encompassing the wall.} Inside of this null cone, the metric is that of an open FRW cosmology
\begin{equation}\label{opends}
ds^2 = -d \tau^2 +a^2(\tau) \left[d \xi^2 + \sinh^2 \xi \ d\Omega_{2}^{2} \right].
\end{equation}
This metric is induced by the embedding
\begin{eqnarray} \label{eq-openslices}
&& X_{0} = a(\tau) \cosh \xi  \\ \nonumber
&& X_{i} = a(\tau) \sinh \xi  \omega_{i}  \\ \nonumber
&& X_{4} = f (\tau),
\end{eqnarray}
where $0 \le \xi < \infty$, $0 < \tau < \infty$, and where $f(\tau)$ solves $f'^2(\tau)=a'^2(\tau)-1$.
If we set $a(\tau)=H_{T}^{-1}\sinh(H_T \tau)$, we have $f(\tau)=H_{T}^{-1}\cosh(H_T \tau)$, and we recover the usual ``open slicing" of dS.

\begin{figure}
\includegraphics[width=7.5cm]{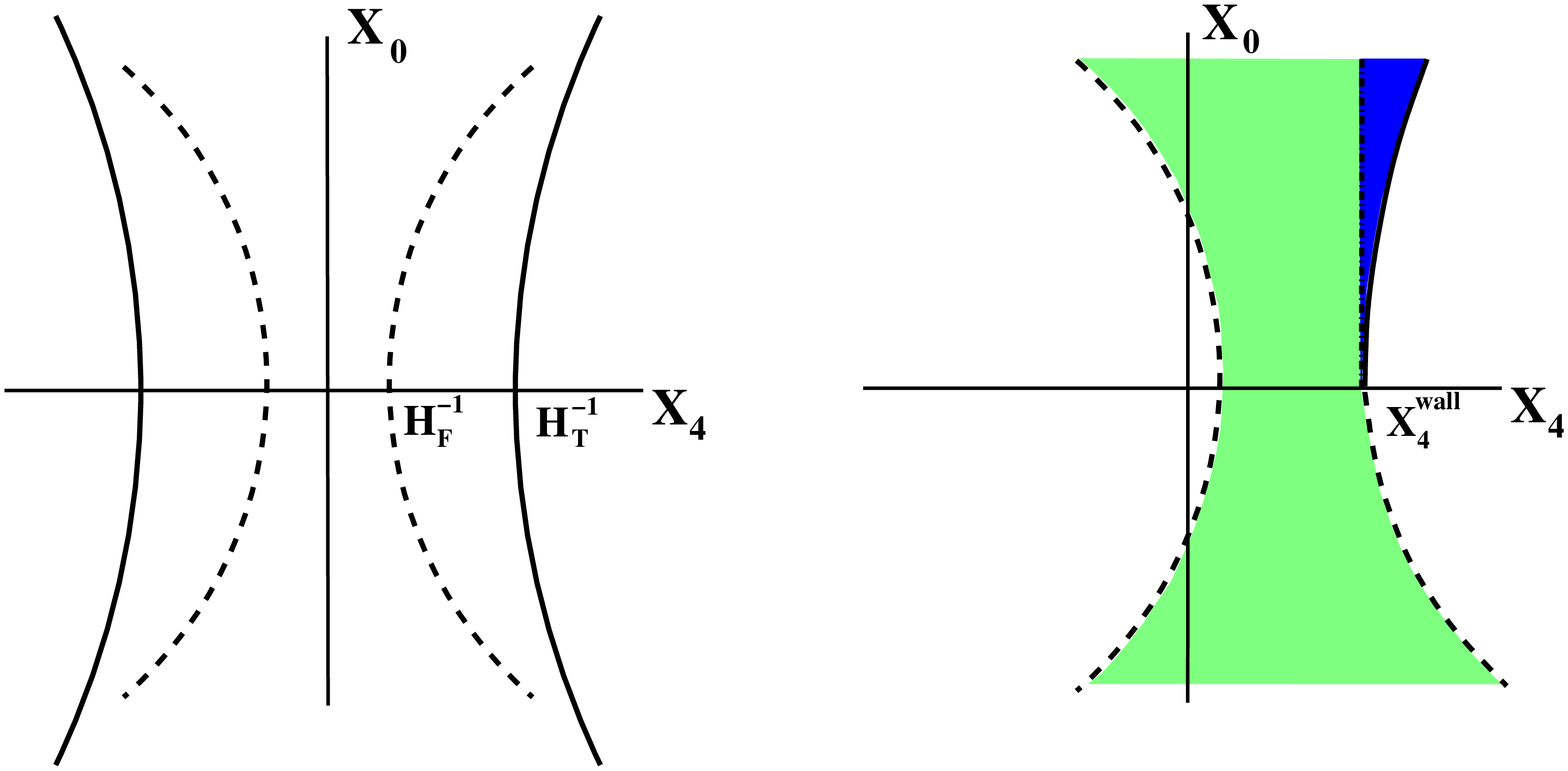}
\caption{On the left is the embedding of two dS spaces of different vacuum energy in 5-D Minkowski space (three dimensions suppressed). The construction obtained by
matching these two hyperboloids along a plane of constant $X_4$, as shown on the right, corresponds to the one-bubble spacetime shown in Fig.~\ref{fig-bubbleanatomy}
in the limit where the bubble interior is pure dS.
The light shaded (green) region represents the false vacuum exterior spacetime, while the dark shaded (blue) region represents the interior spacetime.
  \label{fig-hyperboloids}
}
\end{figure}

Now these two spacetimes can be ``glued together" across the bubble wall.\footnote{Even in the thin-wall limit this is only an approximate solution to the coupled Einstein and scalar field  equations (for the full solution, see eg~\cite{Garriga:1997ef}), corresponding to the limit where the initial bubble radius vanishes.} In the limit where the bubble interior is pure dS, this corresponds to gluing two dS hyperboloids in the embedding space, and breaks the original SO(4,1) symmetry of empty de Sitter space to SO(3,1), since we must choose an axis (here, we choose $X_{4}$) along which to do the pasting. This procedure is shown schematically in Fig.~\ref{fig-hyperboloids}. For a more general interior $a(\tau)$ the picture is similar but with the ``scale" of the hyperboloid varying with $X_4 > X_4^{\rm wall}$.

\begin{figure}
\includegraphics[width=8cm]{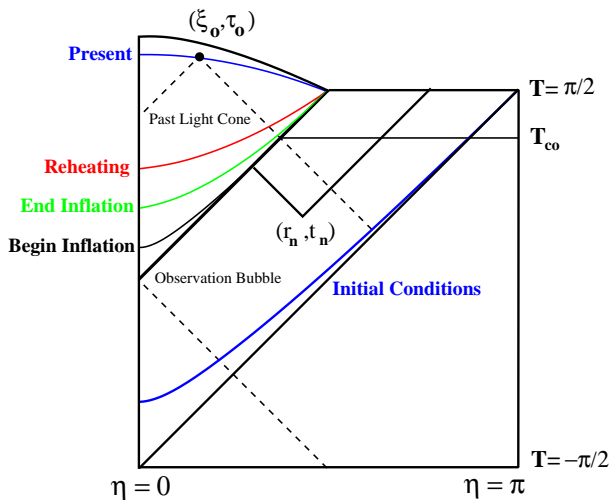}
\caption{The conformal diagram for a bubble universe. We imagine an observer at some position $(\xi_{o}, \tau_{o}, \theta_{o})$ inside of the observation bubble, which is assumed to nucleate at $t=0$ and expand at the speed of light. The foliation of the bubble interior into constant density, negative curvature, hyperbolic slices is indicated by the solid lines. These spacelike slices denote epochs of cosmological evolution in the open FRW cosmology inside of the bubble. The past light cone of the observer is indicated by the dashed lines. There is a postulated no-bubble surface at some time to the past of the nucleation of the observation bubble. Also shown is a ($\theta=\theta_n,\,\phi=\phi_n$ slice of a) colliding bubble that nucleated at some position $(t_n, r_n, \theta_n,\phi_n)$, and intersects the bubble wall within the past light cone of the observer.
  \label{fig-bubbleanatomy}
}
\end{figure}

The basic setup of the problem we wish to consider is shown in Fig.~\ref{fig-bubbleanatomy}, which is the conformal diagram for de Sitter space containing a true vacuum bubble. In this model our observable universe resides within the ``observation bubble." The spacelike slices inside this bubble correspond to surfaces of constant-$\tau$ that, by the homogeneity of the metric Eq.~\ref{opends}, are also surfaces of constant curvature and density. These slices correspond to the various epochs of cosmological evolution inside of the bubble: the beginning of inflation (near the tunnelled-to field value), the end of inflation at the failure of slow-roll, reheating, the recombination epoch, etc., up until the present time.\footnote{We note that it is difficult to construct inflaton potentials (without considerable fine tuning) giving rise to a cosmological evolution inside of the bubble similar to our own.}

If the nucleation rate $\lambda$ (per unit physical 4-volume) of true-vacuum bubbles is small compared to $H_F^{4}$, the observation bubble will be one of infinitely many that form as part of what either is or approaches a ``steady-state" bubble distribution wherein there is a foliation of the background dS in which the bubble distribution is statistically independent of both position and time (see~\cite{Aguirre:2003ck}, and also ~\cite{Vilenkin:1992uf,Aguirre:2001ks}.)  An infinite subset of these will actually collide with the observation bubble.

If we now assume that our bubble experiences what a ``typical" bubble in the steady-state distribution does, then we can follow the strategy of GGV and consider the bubble to exist at $t=0$, model the background as having an initial pure false-vacuum surface at $t=t_0$ (indicated in Fig.~\ref{fig-bubbleanatomy}), then send $t_0\rightarrow -\infty$.  (By doing this, GGV explicitly showed that there is a ``preferred frame" in the model of eternal inflation they treated, which coincides with comoving observers in the ``steady-state" foliation, and is related to the initial false-vacuum surface; observers with different boosts with respect to this frame see bubble collisions at different rates.)

Given an observer at time $\tau_o$ and hyperbolic radius $\xi_o$ inside the bubble, we can define a two-sphere by the intersection of the observer's past lightcone (dashed lines in Fig.~\ref{fig-bubbleanatomy}) with another equal-$\tau$ surface (i.e. corresponding to a portion of the recombination surface or the bubble wall).  The question we now wish to address is: {\bf what is the number of bubbles observed in a given direction $(\theta,\phi)$ with a given angular size on the two-sphere (the observer's ``sky")?} This quantity could provide the basis for a calculation of the impact on the observer's CMB of incoming bubbles that distort the recombination (or reheating, etc.) surface.

In the next section we calculate this quantity under the following assumptions:
\begin{enumerate}
\item{We assume that bubbles start at zero radius and expand at lightspeed at all times.
 We also assume that the bubbles do not back-react, i.e. one bubble will not alter the trajectory of a subsequent bubble. This may be important for directions on the sky hit with multiple bubbles, but requires a careful treatment of bubble collisions and is reserved for future work.}
\item{We assume that no bubbles form within bubbles, and that there are no transitions from true to false vacuum. We comment on the implications of including these features in Sec.~\ref{sec-resimp}}
\item{We assume that structure of the observation bubble is unaffected by the incoming bubbles, and that the observed equal-$\tau$ surface is at $\tau \rightarrow 0$, coinciding with the bubble wall. The first -- rather strong -- assumption is discussed below in Sec.~\ref{sec-resimp}; the second should be reasonable insofar as we are hoping to assess the incoming bubbles' impact on the first few e-foldings of inflation. }
\end{enumerate}

Within this setup, let us examine why it is plausible for a typical observer to have one or more bubble nucleations within their past lightcone.  Because bubbles expand as lightcones and nucleate with some rate $\lambda$ per unit 4-volume, the expected number of bubbles in an observer's past lightcone is just $\lambda V_4$, where $V_4$ is the 4-volume of the exterior spacetime contained in the past light cone of the observer, bounded by the initial value surface, the bubble wall, and the past light cone of the nucleation site of the observation bubble (which enforces the no bubbles-within-bubbles approximation). This 4-volume depends on the position of the observer inside of the bubble and the epoch of observation.

Now, the spatial volume in a coordinate interval $d\xi$ goes as $dV_3 \propto 4\pi \sinh^2\xi d\xi$, thus the volume is exponentially weighted towards large $\xi$. If observers inside of the bubble are uniformly distributed on a given constant-$\tau$ surface, we would expect most of them to exist at large $\xi$.  But as shown by~\cite{Garriga:2006hw}, on any constant-$\tau$ surface, the 4-volume relevant for bubble nucleation diverges for large $\xi$ as $V_4\propto\xi$.  Thus even for a tiny nucleation rate\footnote{We might expect a typical nucleation rate to be of order $\lambda \sim e^{-S_{F}}$, where $S_F$ is the entropy of the exterior de Sitter space.} most observers have a huge 4-volume to their past and should therefore expect bubbles in their past.\footnote{If the interior vacuum energy is much lower than the exterior one, this only increases the 4-volume accessible to the observer.}

We now proceed to calculate the distribution of collisions on our observer's sky. Readers uninterested in the details of this calculation can proceed to Sec.~\ref{sec-resimp} for a summary of the results.

\section{Computations}
\label{sec-computations}

Consider an observer at coordinates $(\xi_o,\tau_o,\theta_o)$ in the observation bubble.  There is nothing breaking the symmetry in $\phi$, so we are free to choose $\phi$=const.
\begin{enumerate}
\item First, we compute the angular scale $\psi$ and direction $\theta_{\rm obs}$
on the sky of the triple-intersection of the observer's past lightcone, the bubble wall (the $\tau \rightarrow 0$ surface), and the wall of a bubble nucleated at some point in the background spacetime.
\item We then find the differential number (Eq.~\ref{difn}) of bubbles of angular size $\psi$ in the direction $\theta_{\rm obs}$ by integrating the volume element for the exterior spacetime over all available nucleation points on a surface of constant $\psi$ and $\theta_{\rm obs}$ and multiplying by the bubble nucleation rate $\lambda$.
\end{enumerate}

Both items can be computed in two different frames that we shall denote the ``unboosted" and the ``boosted" frames.
In the original ``unboosted" frame, where the observer is at $(\xi_o,\tau_o,\theta_o)$, we compute the locations of triple-intersections on the 2-sphere of the observer's sky, then convert these locations to an observed angle $\theta_{\rm obs}$ and angular scale $\psi$ on the sky (see Sec.~\ref{sec-compunboosted} and Appendix~\ref{app-unboosted}).
While this frame is most straightforward, the calculations are much more tractable using a trick suggested by GVV: given the symmetries of dS, a boost in the embedding space changes none of the physical quantities we are interested in (see below for elaboration). Thus we can choose a boost such that the observer lies at $\xi$=0, so that (a) $\theta_{\rm obs}$ coincides with the coordinate angle $\theta_n$ at which the bubble nucleates, and (b) the bubble's angular scale is just given by the angular coordinate separation of the two triple-intersection points. The cost of this simplification is that the initial false-vacuum surface is boosted into a more complicated surface. In the results to follow, we will employ both the boosted and unboosted viewpoints, but will focus on the boosted frame for the calculation of the distribution function.

\subsection{Angles according to the unboosted observer}
\label{sec-compunboosted}

The triple-intersection between the observation bubble, the
colliding bubble, and the past light cone of the observer represent
the set of events that form a boundary to the region on the
observer's sky affected by the collision. Working in a plane of
constant $\phi$, these will correspond to two events, and the angle
between geodesics emanating from these two events and reaching the
observer at $(\tau_o,\xi_o,\theta_o)$ gives the observed angle on
the sky.  In the particular case where the bubble interior is dS
with $H_T=H_F$, Appendix~\ref{app-unboosted} gives the explicit
solution to this problem, although a similar (necessarily more
complicated) procedure can be applied to the more general case.

\begin{figure}
\includegraphics[width=4cm]{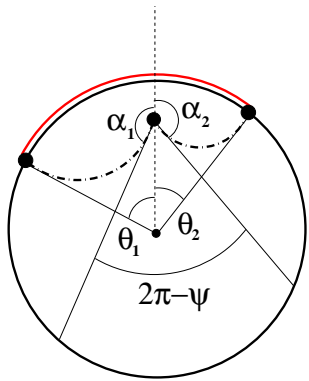}
\caption{
  \label{fig-poincaredisc}
A time lapse picture of the null rays reaching an observer from the boundary of the region affected by a collision event in the Poincar\'e disk representation. The boundaries are located at angles $\theta_{1,2}$ from the center of the disk, and at angles $\alpha_{1,2}$ from the location of the observer. The total angular scale of the collision event as recorded by the observer, which affects the region of the disc indicated by the double lines, is given by $\psi$. }
\end{figure}

Let us visualize this by focusing now on the inside of the
observation bubble which (as discussed in Sec.~\ref{sec-setup}) is
described by an open FRW cosmology. We can use the Poincar\'e disk
representation to describe the hyperbolic equal-$\tau$ surfaces in
this spacetime. Suppressing one of the spatial dimensions, the
metric on a spatial slice of Eq.~\ref{opends}\ becomes
\begin{equation}
ds^2= 4 a(\tau) \frac{dz^2 + z^2 d\theta^2}{(1-z^2)^2}.
\label{eq-poincare}
\end{equation}
Where in terms of the embedding,
\begin{eqnarray}
X_0 & = & a(\tau) \frac{1+z^2}{1-z^2}\\ \nonumber X_1 & = & a(\tau)
\frac{2 z \cos\theta}{1-z^2} \\ \nonumber X_2 & = & a(\tau) \frac{2
z \sin\theta}{1-z^2}
\\ \nonumber X_3 & = & 0 \\ \nonumber X_4 & = & {\rm const.}, \label{eq-poincareembed}
\end{eqnarray}
Since there are collision events that disrupt large angular scales,
we find it useful for visualization purposes to let polar angle
$\theta$ assume also negative values $-\pi<\theta<\pi$ and limit the
range of $\phi$ accordingly. Scaling by $a(\tau)^{-1}$ gives the
disk unit radius, with $z=1$ corresponding to the wall of the
observation bubble, as depicted in Fig.~\ref{fig-poincaredisc}.

This figure shows the time-lapse of a collision event from the perspective of an interior observer on the Poincar\'e disk. The angles $\theta_1$ and $\theta_2$ are the triple-intersection points. The broken lines from these points trace the path of null rays that reach the observer at $(\xi_o, \tau_o, \theta_o=0)$, where we have used the remaining symmetry of the problem to place the observer at $\theta_o=0$.

Analyzing this geometry, the angular position of an intersection from the perspective of an interior observer is given by
\begin{equation}\label{eq-unboostcospsi}
\cos \alpha_{1,2} = \frac{\tanh\xi_o -\cos\theta_{1,2}}{\tanh\xi_o \cos\theta_{1,2}-1}
\end{equation}
Notice that the denominator never vanishes unless $\xi \rightarrow \infty$ (the boundary) where $\cos \alpha=-1$, independent of $\theta$. Using the above results, we conclude that the observer will see a collision as having an angular scale of
\begin{equation}
\psi=\alpha_1-\alpha_2
\end{equation}
where one has to take some care choosing the correct branch of the cosine
function in the process of solving for $\alpha$ using Eq.~\ref{eq-unboostcospsi}, see
Fig.~\ref{fig-poincaredisc}.

Because of the hyperbolic nature of the spatial slices, an observer at large-$\xi_o$ can record an angle $\alpha$ that is very different from $\theta$. To examine this limit, transform to the Euclidean coordinates $(z, \theta)$ on the disc, and expand Eq.~\ref{eq-unboostcospsi} near the boundary at $z=1-\epsilon$
\begin{equation}\label{serf}
\cos\alpha_{1,2}(z,\theta)=-1+\frac{1}{2}\cot^2(\frac{\theta_{1,2}}{2})\epsilon^2+\mathcal{O}(\epsilon^3).
\end{equation}

Accordingly, any given angle $\theta$ gets mapped to $\alpha=\pm \pi$ the closer we approach the boundary $(\epsilon\rightarrow 0)$. On the other hand, regardless of how close to the wall we are, there are always small enough angles $\theta<\epsilon$ that will be mapped by Eq.7 to small hyperbolic angles $\alpha$.

In the first case, choosing the branch of the cosine in Eq.~\ref{eq-unboostcospsi} determines whether the angular size is $\psi \simeq \pi-\pi$ or $\psi \simeq \pi+\pi$. Studying a few examples, it is easy to see that in this limit intersections where $\theta_{1,2}$ have opposite signs get mapped to $\psi \sim 2 \pi$, and intersections where $\theta_{1,2}$ have the same sign get mapped to $\psi \sim 0$.

We will see in the following sections that most of the phase space for bubble nucleation comes from {\it very small angles} $\theta_{\rm obs} \sim 0$, typically yielding one intersection in the upper half and one in the lower half of the disk. In this frame, we also expect the angular scale $|\theta_1-\theta_2|$ to be small, since the majority of colliding bubbles form at very late times, and therefore have a tiny asymptotic comoving size. All of this information taken together suggests that typical collision events will appear to take up {\em either} very large or very small angular scales on the observer's sky, depending on where the observer is situated inside of the bubble.

\subsection{The boosted view}\label{sec-boostedview}
We now go on to discuss the boosted frame. We will again exploit the symmetry of the problem to position the observer at $\theta_o = 0$, and define the following transformation in the embedding space:
\begin{eqnarray}
&& X_{0}' = \gamma \left(X_{0} - \beta X_{1} \right), \\ \nonumber
&& X_{1}' = \gamma \left(X_{1} - \beta X_{0} \right), \\ \nonumber
&& X_{2,3,4}' = X_{2,3,4}.
\label{eq-embeddingboost}
\end{eqnarray}
This is simply a boost in the $X_1$-direction of the embedding space, and respects the SO(3,1) symmetry of the one-bubble spacetime, since it is in a direction perpendicular to the "surface of pasting" described in Sec.~\ref{sec-setup}. If $\gamma = \cosh \xi_{o}$ and $\beta = \tanh \xi_{o}$, the observer at $\xi_o$ is translated to the origin. More generally, in terms of the open coordinates inside of the observation bubble (with arbitrary scale factor), this boost is equivalent to a translation (see Appendix~\ref{app-boost} for an explicit demonstration of this).

Points outside of the observation bubble are also affected by the
boost. We will be particularly concerned with the effects on the
initial value surface at $t_0 \rightarrow \infty$, since this
determines the available 4-volume to the past of our observer. The
boost will push portions of this initial value surface into regions
of the de Sitter manifold not covered by the flat slicing
coordinates (see Eq.~\ref{flatds}). It is therefore useful to employ
the third foliation of dS, into positively curved spatial sections,
which cover the entire manifold. Using a conformal time variable,
these coordinates $(T,\eta,\theta,\phi)$ are defined by:
\begin{eqnarray}
&& X_{0} = H_{F}^{-1} \tan T \\ \nonumber
&& X_{i} = H_{F}^{-1} \frac{\sin\eta}{\cos T} \omega_{i}  \\ \nonumber
&& X_{4} = H_{F}^{-1} \frac{\cos \eta}{\cos T},
\label{eq-closedembedding}
\end{eqnarray}
where $-\pi /2 \leq T \leq \pi / 2$ and $0 < \eta < \pi$, and the $\omega_i$ are the same as in Eq.~\ref{eq-openslices}. This induces the metric
\begin{equation}\label{closedds}
ds^2 = \frac{1}{H_{F}^{2} \cos^2 T} \left[-dT^2 + d\eta^2 + \sin^2 \eta \ d\Omega_{2}^{2} \right].
\end{equation}

The transformation between the boosted and unboosted frames in terms of the global coordinates is given by
\begin{eqnarray}\label{eq-closedboost}
&& \tan \theta' = \frac{\sin \eta \sin \theta}{\gamma \left(\sin \eta \cos \theta - \beta \sin T \right)} \\
&& \tan T' = \gamma \left(\tan T - \beta \frac{\sin \eta \cos \theta}{\cos T} \right)  \\
&& \cos \eta' = \cos T' \frac{\cos \eta}{\cos T}.
\end{eqnarray}

We now apply this transformation to the initial value surface at $t_0 \rightarrow -\infty$. In terms of the embedding coordinates, we can define this (null) surface by $X_{0} + X_{4} = 0$ ($T=\eta - \pi / 2 $), which boosts to
\begin{equation}
X_{0}'+\beta X_{1}' = - \frac{X_{4}'}{\gamma}.
\end{equation}
Substituting with the global coordinates, we arrive at the relation
\begin{equation}\label{boostedinitialvalue}
\sin T' = - \left(\frac{\cos \eta'}{\gamma} + \beta \sin \eta' \cos \theta' \right).
\end{equation}
Henceforward we will drop the prime on the boosted coordinates unless explicitly noted.

\begin{figure}
\includegraphics[width=8.5cm]{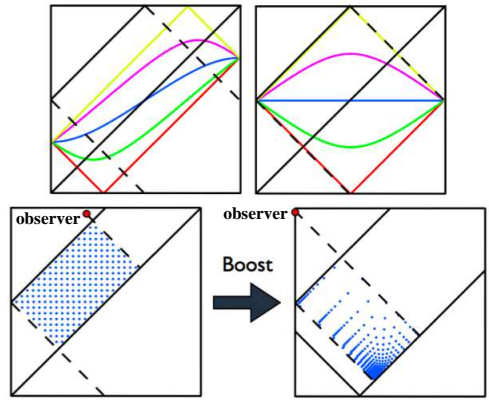}
\caption{
  \label{initialvalue}
The effects of the boost. The top cell shows the boosted initial value surface (at $t_0 \rightarrow -\infty$ in the unboosted frame) for small (left) and large (right) $\xi_o$ for a variety of angles (with the bottom curve (red) corresponding to $\theta = 0$, the top (yellow) corresponding to $\theta = \pi$, and other lines corresponding to intermediate angles at intervals of $\pi/4$). The bottom cell shows the effects of the boost on points in the exterior spacetime on a slice of constant $(\phi,\theta=0)$. Note that even for this very modest boost ($\xi_o = 2$), most of the points are condensed into the wedge created by the past light cone of the nucleation event and the boosted initial value surface.}
\end{figure}

The boosted initial value surface Eq.~\ref{boostedinitialvalue} is a function of the coordinate angle, accounting for the dependence on $\theta_{\rm obs}$ of the past 4-volume for an unboosted observer. This is displayed for a variety of angles on the dS conformal diagrams in the upper cell of Fig.~\ref{initialvalue}.

The effects of the boost on a slice of constant $(\phi, \theta=0)$ in the background spacetime is shown in the lower cell of Fig.~\ref{initialvalue}. Even for this rather modest boost (here we use $\xi_o=2$), it can be seen that most of the points in the unboosted frame are condensed into the wedge between the past light cone of the nucleation event and the boosted initial value surface.

One may be worried that the presence of colliding bubbles, which break the SO(3,1) symmetry of the one-bubble spacetime, invalidates our procedure. In fact, to calculate the quantities we are interested in, we only need a consistent description of the spacetime outside of the colliding bubbles. We assume that the colliding bubbles are null and since SO(3,1) symmetry transformations keep points inside their light cones, it follows that the spacetime outside bubbles is mapped to itself. While it may be true that such transformation may e.g. violate causality {\it inside the colliding bubbles} this effect does not affect the analysis we perform here.

\subsection{Angles according to the boosted observer}\label{sec-boostedangles}

We can now calculate the angular scale of a collision on the boosted observer's sky. To do so, we must confront the non-Euclidean geometry of spatial slices in the global coordinates: constant-$T$ slices are 3-spheres of radius $1/H_F\cos T$. We can visualize a timeslice of bubble evolution by suppressing one dimension, embedding in a 3 dimensional Euclidean space, and scaling the spheres to unit radius. The polar angle on this two-sphere is given by $\eta$ and the azimuthal angle by $\theta$ (recall that we take the range $-\pi < \theta < \pi$).

A bubble wall appears as an evolving circle on the unit 2-sphere. Allowing for arbitrary bubble interiors, and continuing the global coordinate equal time slices ($X_0=$const. in the embedding) into them, a spatial slice is not quite a two sphere, but rather a two sphere with divets and bumps describing the varying curvature of the spacetime inside of the bubbles. For colliding bubbles, these structures -- no  matter how extreme -- are irrelevant, as we will only employ information about the bubble wall.

But the observation bubble requires more care, since we are ultimately interested in a description of collision events from the perspective of an inside observer. Whatever form the embedding of the bubble interior may take, by symmetry, the bubble wall will be a latitude on the background two-sphere. It will have $\eta=T$ (since it nucleates at $T=0$), and span all $\theta$ from $-\pi$ to $\pi$. For $T < \pi/2$ it looks like a circle, with the bubble interior the portion of the sphere bounded by this circle. At $T = \pi/2$ the circle is a great circle and the bubble exterior a hemisphere. If we had chosen a frame in which the observation bubble was formed at some $T_n<0$, then for $T - T_n > \pi/2$ the bubble wall would again become a ``small" circle, with the portion of the sphere bounded by this circle corresponding to the bubble {\em exterior}. By homogeneity of the space a bubble nucleated elsewhere would appear similarly.

\begin{figure}
\includegraphics[width=8cm]{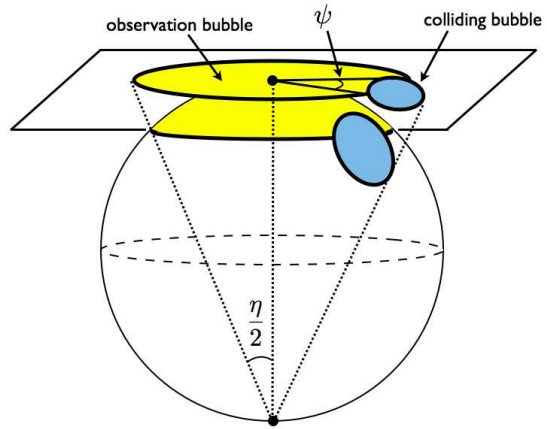}
\caption{
  \label{fig-sphereproj}
A spatial slice in the global foliation of the background de Sitter space, and its stereographic projection. The observation bubble is shaded light (yellow) and the colliding bubble is shaded dark (blue). The angle $\psi$ is indicated in the plane of projection.}
\end{figure}

In the spherically symmetric, open FRW coordinates that describe the interior of the observation bubble, the boosted observer lies at the origin, which coincides with $X_i=0$ in the embedding space. Because of the spherical symmetry of this metric, radial incoming null rays from the bubble wall follow trajectories of constant $\theta$ and $\phi$, and the angle on the sky is identical to the angle we would find if the bubble interior were replaced by a continuation of the background dS. In terms of calculating the observed angle, we can therefore largely ignore the hyperbolic geometry of the bubble interior, and visualize the collision between the observation bubble and an incoming bubble as the intersection of two circles on the $T=$\,const. sphere, as shown in Fig.~\ref{fig-sphereproj}.

In analyzing the geometry it is helpful to perform a stereographic projection onto a plane tangent to the north pole of the two-sphere ($\eta = 0$) as shown in Fig.~\ref{fig-sphereproj}. This projection maps circles on the 2-sphere to circles in the plane, and also preserves angles since the map is conformal.

\begin{figure}
\includegraphics[width=7cm]{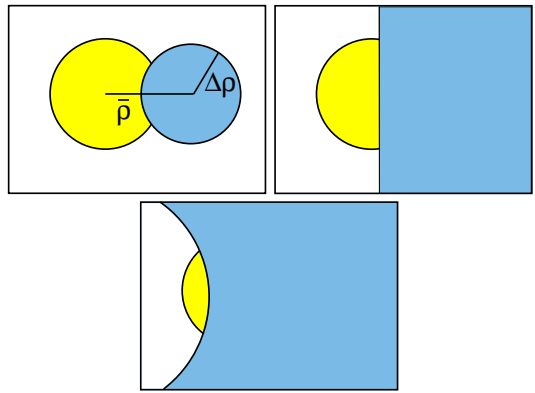}
\caption{
  \label{fig-boostedprojection}
The three cases of bubble intersection in the plane of projection. The top left cell displays the case where the bubble interior does not encompass the south pole of the projected two-sphere, the top right cell displays the case where the bubble wall intersects the south pole, and the lower cell displays the case where the bubble interior includes the south pole.}
\end{figure}

Examining the projection, there are three cases to consider. Colliding bubbles with an interior that does not cut out the south pole appear as filled circles in the projection (upper-left panel of Fig.~\ref{fig-boostedprojection}, where the light (yellow) disc represents the observation bubble and the dark (blue) disc represents the colliding bubble). On the time slice when a bubble wall intersects the south pole, the wall appears as a line in the projection,  bisecting the plane into a region inside, and outside, the bubble (upper-right panel of Fig.~\ref{fig-boostedprojection}). If the bubble interior cuts out the south pole, it projects to a circle whose interior corresponds to the region {\em outside} of the bubble (see the lower panel of Fig.~\ref{fig-boostedprojection}).

Now consider a bubble nucleated at arbitrary coordinates $(T_n,\eta_n,\theta_n)$. Ingoing and outgoing radial null rays from the center of this bubble (corresponding to the location of the bubble wall) obey:
\begin{equation}
\eta=\eta_n\pm (T-T_n)\equiv \eta_n\pm \eta_T.
\end{equation}
We are interested in the projection of this bubble at the global time-slice $T_{\rm co}$ (and bubble coordinate time $\tau_{\rm co}\rightarrow 0$) when the observer's past lightcone intersects the observation bubble wall (see Fig.~\ref{fig-bubbleanatomy}). If we follow the past lightcone of the observer we find
\begin{equation}
\xi=\int_\tau^{\tau_o}d\tau/a(\tau).
\end{equation}
To determine $T_{\rm co}$, a valid junction between the interior and exterior spacetimes requires that the physical radius of two-spheres (the coefficients of $d\Omega_2$ in Eq.~\ref{opends} and~\ref{closedds}) at the location of the wall match, and gives
\begin{equation}\label{Tco}
T_{\rm co}=\arctan \left[ H_F \lim_{\tau \rightarrow 0} a(\tau) \sinh \left( \int_{\tau}^{\tau_{\rm o}}d\tau/a(\tau) \right) \right].
\end{equation}
In the case where the interior is pure dS (where $a(\tau)=H_T^{-1}\sinh H_T\tau$), this works out to $T_{\rm co}=\arctan [(H_F/H_T)\tanh (H_T \tau_o/2)]$. As we send $\tau_o \rightarrow \infty$, it can be seen that this ranges between $T_{\rm co}=\pi / 4$ for $H_{T} = H_{F}$ and $T_{\rm co} = \pi / 2$ for $H_{T} \ll H_{F}$.

Viewed in the projected plane using polar coordinates $(\rho,\phi_{\rm proj}$), the incoming bubble has a center at $\bar\rho = (\rho_2+\rho_1)/2$, and a radius $\Delta\rho = (\rho_2-\rho_1)/2$ as shown in the upper left panel of Fig.~\ref{fig-boostedprojection}. Then, since the projection of an arbitrary point gives $\rho=2\tan \eta/2$ (this can be seen by analyzing the geometry of Fig.~\ref{fig-sphereproj}), we can work out:
\begin{equation}
\bar\rho=\frac{2 \sin\eta_n}{\cos\eta_n+\cos\eta_T},\ \
\Delta\rho=\frac{2 \sin\eta_T}{\cos\eta_n+\cos\eta_T}.
\end{equation}
Finally, on the plane we can find the angle $\psi$ between the two radial null rays that come to the observer from the two intersection points, which is given by:
\begin{equation}
\cos \left( \frac{\psi}{2} \right) = -\cot \eta_n \cot T_{\rm co} + \frac{\cos (T_n - T_{\rm co})}{\sin \eta_n \sin T_{\rm co}}.
\label{eq-psionsky}
\end{equation}
At $\xi=\eta=0$, observers at rest in the open and closed coordinates are in the same frame, so $\psi$ is the actual angular scale on the sky of the bubble's ``sphere of influence", as seen by the observer.

\begin{figure}
\includegraphics[width=8.5cm]{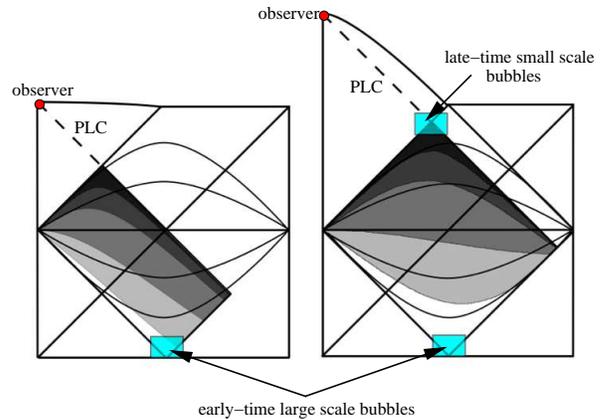}
\caption{
  \label{fig-angles}
The foliation of the exterior de Sitter space into surfaces of constant $\psi$ for junctions with $H_{T} \sim H_{F}$ (left) and $H_{T} \ll H_{F}$ (right). Dark regions correspond to small $\psi$ and light regions correspond to large $\psi$. Superimposed on this picture is the boosted initial value surface for various $\theta_n$ in the limit of large-$\xi_o$.}
\end{figure}

We can now foliate the background spacetime into surfaces of constant $\psi$, as shown in Fig.~\ref{fig-angles}. From the symmetries of the boosted frame, this foliation is independent of $\theta$ and $\phi$ (although the angular dependence of the boosted initial value surface will play an important role in defining the statistical distribution of collisions). This provides a map between the nucleation site of a colliding bubble and the observed angular scale of the collision. The number of collisions of a given angular scale can be found by examining how the exterior four-volume is distributed in the causal past of the observer.

In the $\xi_o \rightarrow \infty$ limit, there is a {\em divergent} 4-volume containing nucleation sites that correspond to $\psi \sim 2 \pi$ and $\theta_n \simeq 0$ (in the corner near past null infinity enclosed by the shaded boxes of Fig.~\ref{fig-angles}, the left panel of which shows the $H_{T} \sim H_{F}$ case). Considering the time evolution of an observer starting from $\tau \simeq 0$, most of the 4-volume in this region will come into the observer's past light cone at very early times. The observer will therefore see new bubble collisions at a rate that is very high at first (formally divergent as $\xi \rightarrow \infty$), and decreases with time\footnote{Surfaces of constant $\xi$ are nearly null at early times, so this effect can be viewed as due to time dilation in the boosted frame.}.

In the limit where $H_{T} \ll H_{F}$, for all $\xi_o$, there is also a very large 4-volume containing nucleation sites that correspond to $\psi \sim 0$ (in the corner near future null infinity enclosed by the shaded box), though the observer will not have access to these collisions until late times. In this late-time limit (and even for $\xi_o \rightarrow \infty$), the boosted initial value surface cuts into the relevant phase space only when $\theta_{\rm obs} \sim \pi$, so the distribution is nearly isotropic.

Assembling this information, we predict that the distribution function has two potentially large peaks: one at $\psi \sim 2 \pi$ and $\theta_n =0$, for large $\xi_o$, and one at
 $\psi \sim 0$ and all angles, for large $\tau_o$; both are in complete agreement with the analysis of the unboosted frame. Collisions with $\psi \sim 2 \pi$ are recorded at very early observation times, while those with $\psi \sim 0$ are recorded at very late observation times. We now directly confirm these predictions by explicitly calculating the distribution function in the boosted frame.

\subsection{Angular distribution function}\label{sec-distfunc}

We now calculate $\frac{dN}{d\psi d\cos \theta_{\rm obs}  d\phi_{\rm obs}}$,  the differential number of bubbles with an observed angular scale $\psi$ in a direction on the sky given by $(\theta_{\rm obs} , \phi_{\rm obs})$. In Sec.~\ref{sec-boostedangles} we found a mapping (Eq.~\ref{eq-psionsky}) between the position at which a colliding bubble nucleates and the observed angular scale $\psi$ as seen by an observer situated at the origin (for which $\theta_{\rm obs}=\theta_n, \phi_{\rm obs}=\phi_n)$. We can therefore calculate the distribution function by determining the density of nucleation events on surfaces of constant $\psi$ and $\theta_n$. (The symmetry in $\phi$ implies that the distribution is independent of $\phi_n$.)

The differential number of bubbles nucleating in a parcel of 4-volume somewhere to the past of the observation bubble is:
\begin{equation}\label{eq-volumeelement}
dN = \lambda dV_4 = \lambda H_{F}^{-4} \frac{\sin^2 \eta_n}{\cos^4 T_n} dT_n d \eta_n d(\cos \theta_n) d\phi_n.
\end{equation}

A more complete analysis would include the probability that a given nucleation site is not already inside of a bubble. Under our assumption that bubble walls are null, this probability is given by $f_{\rm out} = e^{-\lambda V_{4}^{\rm past}(\eta_n, T_n, \theta_n)}$~\cite{Guth:1981uk}, where $V_{4}^{\rm past} (\eta_n, T_n, \theta_n)$ is the 4-volume to the past of a given nucleation point. Consider some parcel of 4-volume from which bubbles might nucleate. At late times, in the unboosted frame, a straightforward calculation shows that the 4-volume to the past of any point is proportional to $t$, the flat slicing time. This yields a differential number of nucleated bubbles:
\begin{equation}
\frac{dN}{dt dr d(\cos \theta) d \phi} = \lambda r^2 e^{(3 -\lambda H^{-4}) H t}  \simeq \lambda r^2 e^{3 H t},
\end{equation}
where we have used the fact that in any model of eternal inflation $\lambda H^{-4} \ll 1$.
The total number of bubbles is found by integrating, and it can be seen (essentially for the same reason that inflation is eternal in these models) that including $f_{\rm out}$ only minutely affects both the differential and total bubble counts. We will therefore neglect this correction in our calculation.

Returning to Eq.~\ref{eq-volumeelement}, changing variables from $T_n$ to $\psi$ using Eq.~\ref{eq-psionsky}, and integrating $\eta_n$ at constant $\psi(\eta_n,T_n)$, we obtain the distribution function:
\begin{widetext}
\begin{equation} \label{dndpsi}
 \frac{dN}{d\psi d(\cos\theta_{\rm obs}) d\phi_{\rm obs})} = \frac{dN}{d\psi d(\cos \theta_n) d\phi_n} = \lambda H_{F}^{-4} \left[ \int_{0}^{\eta_{\rm max} (\xi_o, \psi, \theta_n)} d \eta_n \frac{\sin^2 \eta_n}{\cos^4 (T_n(\psi,\eta_n, T_{\rm co})) } \left| \frac{\partial T_n(\psi,\eta_n, T_{\rm co})}{\partial \psi} \right| \right],
\end{equation}
with the Jacobian given by
\begin{equation}
\left| \frac{\partial T_n(\psi,\eta_n, T_{\rm co})}{\partial \psi} \right| = \frac{1}{2} \sin \eta_n \sin T_{\rm co} \sin \left(\frac{\psi}{2} \right) \left[1 - \left( \cos \left(\frac{\psi}{2} \right) + \cot \eta_n \cot T_{\rm co}\right)^2 \sin^2 \eta_n \sin^2 T_{\rm co} \right]^{-1/2}.
\end{equation}
\end{widetext}

The lower limit of integration at $\eta_n=0$ can be understood by
tracing the surfaces of constant $\psi$ in Fig.~\ref{fig-angles} and
also by noting that for all $\psi$ and $T_{\rm co}$,
Eq.~\ref{eq-psionsky} yields $T_n(\psi, \eta_n = 0, T_{\rm co}) =
0$. The upper limit of of integration, $\eta_{\rm max}(\xi_o, \psi,
\theta_n)$, is found by determining the intersection of the surfaces
of constant-$\psi$ with the boosted initial value surface; this
intersection depends on $\theta_n$ and $\xi_o$ (due to the boosted
initial value surface Eq.~\ref{initialvalue}), reflecting the
dependence of the past 4-volume on the position of the observer.

The properties of the observation bubble enter this calculation through the determination of $T_{\rm co}$ via Eq.~\ref{Tco}. Recall that for late-time observers ($\tau_o \rightarrow \infty$), $T_{\rm co}$ can range from $\frac{\pi}{4}$ for $H_T = H_F$ to $\frac{\pi}{2}$ for $H_T \ll H_F$.

\begin{figure}
\includegraphics[width=7cm]{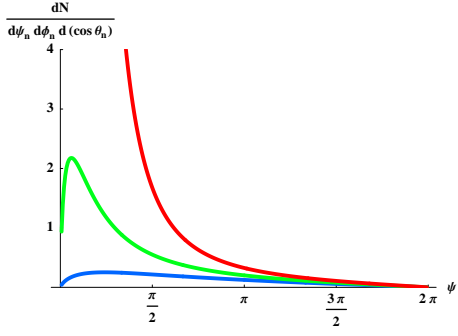}
\caption{The distribution function Eq.~\ref{dndpsi} for an observer
at $\xi_o = 0$ with (from the blue curve on the bottom to the red
curve on top) $T_{\rm co}=\frac{\pi}{4},\frac{3
\pi}{8},\frac{\pi}{2}$ (corresponding to a varying $H_T$), factoring
out the overall scale $\lambda H_{F}^{-4}$. (This factor will in
general be astronomically small, but we choose this convention to
more clearly display the functional behavior of the distribution
function.) This function is independent of $\theta_n$ for this
observer. As $T_{\rm co} \rightarrow \pi/2$ ($H_T / H_{F}
\rightarrow 0$), a divergent peak around $\psi=0$ develops.
  \label{fig-dnunboosted}
}
\end{figure}

We first examine the behavior of the distribution function Eq.~\ref{dndpsi} for an observer at the origin, $\xi_o=0$. In this limit, the distribution is isotropic, and based upon the discussion surrounding Fig.~\ref{fig-angles}, we expect it to have a large peak around  $\psi=0$ as $T_{\rm co} \rightarrow \pi / 2$ ($H_T / H_{F} \rightarrow 0$ and $\tau_o \rightarrow \infty$). Integrating Eq.~\ref{dndpsi}, we see in Fig.~\ref{fig-dnunboosted} that this behavior is indeed observed. For fixed $H_T / H_{F}$, the amplitude of the distribution function approaches a constant maximum value as $\tau_o \rightarrow \infty$ ($T_{\rm co}$ approaches its maximum). We will see in the next section that the total number of observable collisions at late times is bounded, reflecting the behavior of the distribution function.

\begin{figure}
\includegraphics[width=7cm]{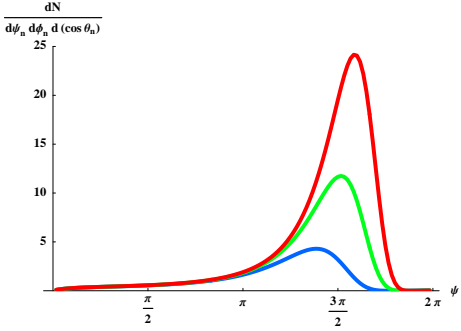}
\caption{The distribution function Eq.~\ref{dndpsi} for an observer at $\xi_o = 25$, with $T_{\rm co} = \frac{\pi}{4}$, for $\theta = \frac{\pi}{10}, \frac{\pi}{15}, \frac{\pi}{20}$, factoring out the overall scale  $\lambda H_{F}^{-4}$. As $\theta_n \rightarrow 0$, the position of the peak shifts to larger $\psi$, and increases in amplitude, displaying the predicted anisotropic peak about large angular scales.
  \label{fig-dndpsi1}
}
\end{figure}

From the analysis of the boosted initial value surface in Sec.~\ref{sec-boostedview}, we predicted that in the limit of large-$\xi_o$, the distribution function Eq.~\ref{dndpsi} should be anisotropic, peaking around $\theta_n = 0$. Fig.~\ref{fig-dndpsi1} shows a number of constant-$(\theta_n, \phi_n)$ slices through the distribution function for $T_{\rm co} = \frac{\pi}{4}$ and $\xi_o = 25$, where we see that this behavior is indeed present. The peak at large $\psi$, which was predicted to arise based upon the analysis in both the unboosted (Sec.~\ref{sec-compunboosted}) and boosted frames (Sec.~\ref{sec-boostedangles}), is present in this example as well. Finally, we observe that as $\theta_n \rightarrow 0$, the distribution peaks at progressively larger $\psi$. This feature can be predicted from Fig.~\ref{fig-angles} by noting that as $\theta_n \rightarrow 0$, an increasing fraction of the 4-volume above the boosted initial value surface corresponds to nucleation sites that produce a large $\psi$ (the shaded box near past null infinity in Fig.~\ref{fig-angles}).

\begin{figure}
\includegraphics[width=7cm]{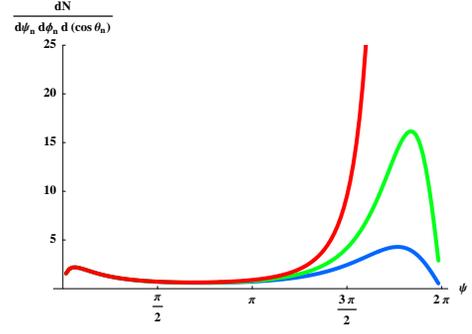}
\caption{The distribution function Eq.~\ref{dndpsi} with $\theta_n = 0$ and $\tau_{\rm co} = \frac{3\pi}{8}$ for $\xi_o = (1.5, 2, 100)$, factoring out the overall scale  $\lambda H_{F}^{-4}$. As $\xi_o$ gets large, the peak near $\psi \sim 2 \pi$ grows, while the peak near $\psi \sim 0$ remains of constant amplitude.
  \label{fig-dndpsi2}
}
\end{figure}

Focusing on a slice through the distribution function with $(\theta_n=0, \phi_n={\rm const.})$ -- for which the amplitude is largest -- we can study the effects of varying $T_{\rm co}$ and $\xi_o$. Fig.~\ref{fig-dndpsi2} shows the distribution function for fixed $\theta_n = 0$ and $T_{\rm co}=\frac{3 \pi}{8}$ with varying $\xi_o$. As $\xi_o$ increases, the amplitude of the peak at large $\psi$ increases, while the peak at small $\psi$ remains unaffected. This can be understood from Figs.~\ref{initialvalue} and \ref{fig-angles} by recognizing that as $\xi_o$ grows, the phase space near past null infinity -- corresponding to nucleation points producing $\psi \sim 2\pi$ -- grows, while the phase space near the intersection of the past light cone and the observation bubble wall -- corresponding to nucleation points producing $\psi \sim 0$ -- remains constant.

\begin{figure}
\includegraphics[width=7cm]{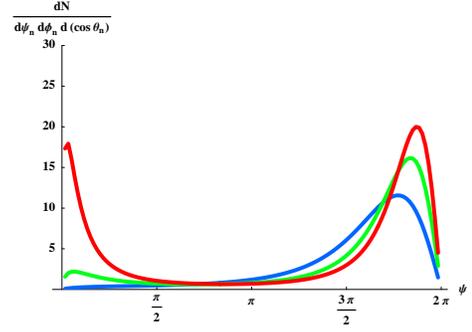}
\caption{The distribution function Eq.~\ref{dndpsi} with $\theta_n = 0$ and $\xi_o = 2$ for  $T_{\rm co} = (\frac{\pi}{4}, \frac{3 \pi}{8}, \frac{7 \pi}{16})$, factoring out the overall scale  $\lambda H_{F}^{-4}$. As $T_{\rm co}$ grows, the bimodality of the distribution becomes more and more pronounced. Both the peak about $\psi \simeq 0$ and $\psi \simeq 2 \pi$ grow, with the growth of the $\psi \simeq 0$ peak eventually overtaking the growth of the $\psi \simeq 2 \pi$ peak. The position of the peaks shift as well, with one peak approaching $\psi = 0$ and the other $\psi = 2 \pi$ as $T_{\rm co} \rightarrow \frac{\pi}{2}$.
  \label{fig-dndpsi3}
}
\end{figure}

Finally, Fig.~\ref{fig-dndpsi3} shows the evolution of the distribution function produced by fixing $\theta_n = 0$ and position $\xi_o = 2$ and increasing $T_{\rm co}$ (corresponding to the actual time-evolution of the distribution function seen by this observer). Here, the bimodality of the distribution becomes apparent. Based on Fig.~\ref{fig-angles}, we determined that bubbles with large angular scales form at early (open slicing) observation times, and bubbles with small angular scales form at late times. This can be seen in the distribution function of Fig.~\ref{fig-dndpsi3}. As $T_{\rm co}$ increases, the peak near $\psi \simeq 0$ becomes more and more pronounced,  overtaking the amplitude of the $\psi \simeq 2\pi$ peak, whose growth eventually stagnates. The positions of the peaks also shift, moving towards $\psi = 0$ and $\psi = 2 \pi$, respectively, as $T_{\rm co}$ increases.

\subsection{Behavior of the distribution near $\psi \simeq 2\pi$ and $\psi \simeq 0$}\label{sec-divergence}

Since the distribution function (as displayed in the figures) is multiplied by $\lambda H_{F}^{-4} \ll 1$, it must have a very large amplitude for our hypothetical observer to hope to see any collisions. We have seen that the distribution function is largest for $\psi \simeq 2 \pi$ (corresponding to collisions occurring at small $\tau$) in the large-$\xi_o$, small-$\theta_n$ limit  as well as for $\psi \simeq 0$ (corresponding to collisions occurring at large $\tau$) in the limit where $H_T \ll H_F$. The origin of these peaks was discussed in Sec.~\ref{sec-boostedangles}, but now we assess them quantitatively.

\subsubsection{The peak at $\psi \sim 0$}

The total number of late-time collisions can be found by evaluating $\lambda$ times the 4-volume $V_{4}^{\psi \sim 0}$ in the exterior spacetime corresponding to small angles. Assuming that the bubble interior and exterior are pure dS and taking the limit of large $\tau_o$ with $H_T \ll H_F$, we obtain
\begin{equation}\label{eq-nbound}
N^{\psi\sim0} = \frac{4 \pi \lambda}{3 H_{T}^{2} H_{F}^{2}} \tanh^2 \left( \frac{H_T \tau_o}{2} \right) + \mathcal{O}\left(\log \frac{H_F}{H_T} \right).
\end{equation}
For fixed $H_T$ this approaches a fixed number as $\tau_o\rightarrow \infty$, but this number can be arbitrarily large if $H_T\rightarrow 0$. We see also that for $N^{\psi\sim0}  \agt 0$, we require both $H_{T} < \lambda^{1/2} H_{F}^{-1}$, and $\tau_o \agt H_F \lambda^{-1/2}$.

\begin{figure}
\includegraphics[width=7cm]{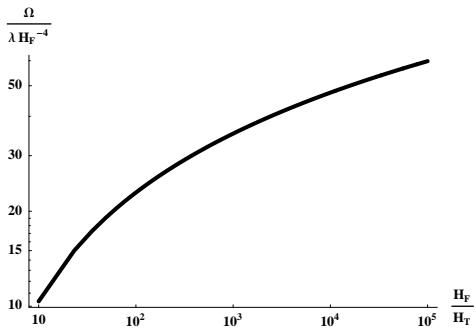}
\caption{A log-log plot (calculated numerically) of the total angular area on the sky taken up by late-time collisions with $\psi \simeq 0$.
  \label{HfHt}
}
\end{figure}

The angular scale of late-time collisions decreases with $\tau_o$, as exhibited by Fig.~\ref{fig-dndpsi3}; one might then ask what total angular area on the sky is affected. This can be found by evaluating:
\begin{equation}
\Omega = \lambda \int dV_4 \psi^2
\end{equation}
over the volume outside of the observation bubble available for the nucleation of colliding bubbles, where $\psi$ is a function of the exterior spacetime coordinates as in Eq.~\ref{eq-psionsky}.  As it turns out, the decrease in angular scale nearly cancels the growth in  $N^{\psi\sim0}$, so while the latter scales as $(H_F/H_T)^2$, the maximal sky fraction is nearly logarithmic in $H_F/H_T$, as shown in Fig.~\ref{HfHt}. Since $\lambda H_{F}^{-4} \ll 1$, the total angular area is very small unless $H_T$ is essentially zero (and $\tau_o$ absurdly large); thus for any realistic scenario the bubble distribution should be considered a set of point sources with infinitesimal total solid angle.

\subsubsection{The peak at $\psi \sim 2\pi$}

Let us now consider the large-$\xi_o$, small-$\theta_n$ limit. To do
so, we take $\psi = 2\pi -\epsilon$ with $\epsilon \ll 1$ and look
at $T_{\rm co} = \pi / 4$ (the amplitude of the peak would only be
larger if we were to take $T_{\rm co} > \pi / 4$, so this gives a
lower bound). Keeping terms to first order in $\epsilon$, we can
simplify the various objects in Eq.~\ref{dndpsi} immensely: $T_n$
along constant $\psi$ surfaces is given approximately by $T_{n} = -
\eta_{n}$, and the Jacobian reduces to
\begin{equation}
\left| \frac{\partial T_n(\psi,\eta_n)}{\partial \psi} \right| = \frac{\epsilon}{4} \frac{\sin \eta_n}{\sqrt{1+\sin (2 \eta_n)}}
\end{equation}
yielding a distribution
\begin{eqnarray}\label{eq-dnlimit}
\frac{dN}{d\psi d\phi_n d(\cos \theta_n)} &=& \frac{\lambda H_{F}^{-4} \epsilon}{4} \times \\ \nonumber
&& \int_{0}^{\eta_{\rm max}} \frac{(\tan \eta_n)^{3}}{\cos \eta_n \sqrt{1+\sin(2 \eta_n)}} d\eta_n.
\end{eqnarray}

In the limiting case under discussion, we can solve for $\eta_{\rm max}$ from the simplified form of the initial value surface (obtained from Eq.~\ref{boostedinitialvalue})
\begin{equation}
\sin \eta_{\rm max} = \frac{\cos \eta_{\rm max}}{\gamma} + \beta \sin \eta_{\rm max}
\end{equation}
yielding
\begin{equation}
\eta_{\rm max} = \sec^{-1} \left(e^{\xi_o} \sqrt{1+e^{-2 \xi_o}} \right),
\end{equation}
where we have not yet taken $\xi_o$ large. Integrating Eq.~\ref{eq-dnlimit}, substituting with $\eta_{\rm max}$, and taking $\xi_o \gg 1$, we obtain:
\begin{equation}\label{eq-2pidivergence}
\frac{dN}{d\psi d\phi_n d(\cos \theta_n)} = \frac{\lambda H_{F}^{-4} \epsilon}{12} e^{3 \xi_o},
\end{equation}
which diverges as $\xi_o \rightarrow \infty$.

Integrating the distribution function over the neighborhood of $\psi
\sim 2 \pi$ and small $\theta_n$ would yield the total number of
observed early-time collisions, which is given by $N^{\psi \sim 2
\pi} \simeq 4 \pi \xi_o \lambda H_{F}^{-4}$ \cite{Garriga:2006hw}.
Since each of these collision events can in principle affect an
angular scale of order $\psi \simeq 2 \pi$, only a vanishing
fraction of the total angular area on the sky remains unaffected in
the $\xi_o \rightarrow \infty$ limit (unlike the long-time limit of
late-time collisions discussed above).

\section{Summary of results and implications}\label{sec-resimp}

\subsection{Properties of the distribution function}

Given an observer at some point in their bubble defined by $(\tau_o,\xi_o,\theta_o=0)$, we have calculated the expected number, angular size, and direction $(\theta_{\rm obs},\phi_{\rm obs})$ of regions on the sky affected by bubble collisions, under the assumption that those collisions merely perturb the observation bubble.

Three key features of this distribution $dN/d\psi d(\cos\theta_{\rm obs}) d\phi_{\rm obs}$ are:
\begin{itemize}
\item For observers at $\xi_o \neq 0$ inside bubbles with $H_T \ll H_F$, the distribution is bimodal, with peaks at $\psi \simeq 0$ and $\psi \simeq 2 \pi$ forming at late and early observation times respectively.
\item For early-time collisions with $\psi \simeq 2 \pi$, the distribution is strongly anisotropic as $\xi_o \rightarrow \infty$, with the overwhelming majority of collision events originating from $\theta_{\rm obs} \simeq 0$, while the distribution of collision events with $\psi \simeq 0$ becomes isotropic at late-times.
\item For a given $H_{T}$, $H_F$, and $\tau_o$, the peak at $\psi \simeq 2\pi$ diverges as $\exp(3 \xi_o)$; the peak at $\psi \simeq 0$ has fixed amplitude, with the total number of such collisions bounded by $N^{\psi \sim 0} \alt \lambda H_{T}^{-2} H_{F}^{-2}$.
\end{itemize}

Although different observers see qualitatively different bubble distributions, we can focus on two key classes: those at large $\xi_o$ and those at very late times $\tau_o$.

Because the bubble interior is naturally foliated into a set of homogeneous spaces that accord no particular preference to $\xi_o=0$, we might imagine observers distributed uniformly over these spaces. In this case (as argued in Sec.~\ref{sec-setup}) a ``typical" observer would be at large $\xi_o$, and have causal access to a large number of collision events (as long as $\xi_o \agt H_{F}^{4} \lambda^{-1}$). If such collisions are Compatible (with our observations), we should therefore expect that they exist to our past.

At very late times, observers at any position $\xi_o$ will have access to nearly the same  distribution of collisions. We have seen that such an observer would typically record the first collision at exponentially late times (of order $\tau_o \sim \lambda^{-1/2} H_{F}$), with tiny angular scale. Thereafter, the number of collisions would grow to asymptotically approach $\sim H_{T}^{-2} H_{F}^{-2}$, and the distribution would become nearly isotropic.  Note that this analysis is relevant to the suggestion by
~\cite{Maloney:un,Shenker:06} that an observer residing at $\xi_o=0$ inside of a bubble with $H_T = 0$ (the ``census taker" of~\cite{Shenker:06}) could be used to define a measure over the pocket universes in eternal inflation; it may also be relevant for evaluating the quantum-gravitational degrees of freedom of an eternally-inflating de Sitter space~\cite{Arkani-Hamed:2007ky}. In terms of {\em our} observations,
if we fix $H_T$ to be the vacuum energy we currently observe, and $\tau_o \sim H_T^{-1}$, late-time, small angular scale collisions could be observable if $\lambda H_{F}^{-4} \agt 10^{-100}$. While perhaps an atypically large tunneling rate, this is well within the limit $\lambda H_F^{-4} << 1$ required for eternal inflation in our parent vacuum.

Because all observers might potentially `see' bubbles at late times (for sufficiently large $\lambda$), and essentially (except for a set of measure zero) all should `see' collisions at early times, it is interesting to ask what potential observational effects might exist.

\subsection{A classification of collision events}

Unfortunately, assessing any potentially observational effects of this scenario requires a good understanding of the outcome of bubble collisions under a variety of circumstances, which is presently lacking.  As a preliminary step, we can qualitatively survey the general types of collision events that might occur in a universe undergoing false-vacuum eternal inflation; after this we return to what these collision types could imply observationally.

\begin{figure}
\includegraphics[width=6cm]{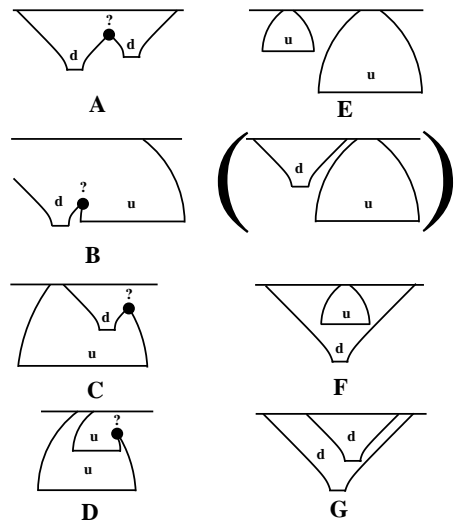}
\caption{A general set of situations which might involve collisions between two bubbles in an eternally inflating spacetime. Each cell represents a region near future null infinity (horizontal solid line) of an eternally inflating background dS. True vacuum bubbles form at very small radius and expand, while false vacuum bubbles form larger than the exterior horizon size, and contract. Collisions are denoted by filled circles, with the uncertainty of the post-collision spacetime indicated by a question mark.
  \label{fig-gencoll}
}
\end{figure}

Each cell of Fig.~\ref{fig-gencoll} depicts two bubbles near future
null infinity in the eternally inflating background dS. Cell A
depicts the situation considered thus far, of two colliding true
vacuum bubbles (``downward-bubbles" for present purposes). Others
show also transitions {\em upward} from the false vacuum
(``upward-bubbles"); the structure of such bubbles is very
different: they collapse due to the inward pressure
gradient~\cite{Lee:1987qc}, so if they contain a finite region of
future null infinity, then they must form with super-(exterior)
horizon size\footnote{Or form on the other side of an Einstein-Rosen
bridge.} (e.g.,~\cite{Aguirre:2005sv}).

The first column (A-D) shows situations where bubbles actually
collide; the right-hand column (E-G) shows cases in which the
dynamics of the bubble walls prevent a collision from occurring (we
display these solutions to illustrate that there are cases in which
pairs of bubbles nucleated rather close to one another will not
collide). Concentrating on the first column, cell A shows the
collision between two downward-bubbles (which may or may not be of
the same vacuum energy). Downward-bubbles can also collide with
upward bubbles (cell B), but because the latter accelerate inward,
and have strongly suppressed formation rates relative to
downward-bubbles, such collisions should be extremely rare.
Collisions of type C, between nested bubbles, occur if a
downward-bubble quickly nucleates within an upward-bubble, our
observation bubble is unlikely to be such an early bubble --
infinitely many others will form later within the same false-vacuum
bubble. Finally, nested upward-bubbles may collide (cell D), but
only very rarely.

This general survey of two-bubble collision events, indicates that the focus on situation A alone is quite justified: all other possible collision events should be negligibly rare.

Determining the detailed aftermath of a collision event between two vacuum bubbles of arbitrary vacuum energy is a very complicated problem, most likely involving numerical relativity. Previous numerical and analytic studies have treated cases where the vacuum energy inside both bubbles vanishes~\cite{Hawking:1982ga,Bousso:2006ge}, cases where both bubbles have negative vacuum energy~\cite{Blanco-Pillado:2003hq}, and cases where a zero and negative vacuum energy bubble collide~\cite{Freivogel:2007fx}.

In the absence of detailed computations, but based on these studies, we can outline a few generic possibilities.  For collisions between bubbles of the same vacuum, the disturbed intersection region might radiate away much of the wall's energy, then be smoothed out by subsequent inflation.  For bubbles of different vacuum field value, wall energy may still radiate away (as demonstrated in~\cite{Hawking:1982ga,Blanco-Pillado:2003hq}), but a domain wall must remain, and would presumably accelerate into the bubble of higher vacuum energy.

In terms of the effect on an observation bubble, it would seem that collisions resulting primarily in a domain wall accelerating away from an observer are likely to be ``Compatible" (in the terminology of Sec.~\ref{sec-intro}) over a significant part of the collision's future. Even if considerable energy is released, it will be red-shifted by the epoch of inflation within the bubble, perhaps resulting in only a minor perturbation of the interior cosmology. On the other hand, a domain wall accelerating {\em towards} the observer will almost certainly be catastrophic (and hence not Compatible). In between, bubbles of the same vacuum (where there is no domain wall), or collisions resulting in a timelike domain wall (as in~\cite{Freivogel:2007fx}), may or may not be Compatible (for all or just a portion of the causal future of the collision) depending on the details of the collision.

Returning to Fig.~\ref{fig-gencoll}, cells A-C depict collision events potentially relevant to the observation bubble. In each case, if the vacuum energy of the observation bubble is lower than the vacuum energy of both the background dS and the colliding bubble, it seems likely that the collision is Compatible over most of its future. A or C could alternatively be fatal if the incoming bubble (in cell A) or the background space (in cell C) are at lower vacuum energy than the observation bubble. However, the finer details will need to be studied to provide a definitive classification of these collision events and to what degree they satisfy the Compatibility condition.

\subsection{Observational implications}

What does all of this mean for making predictions starting from a fundamental theory that drives eternal inflation? The above discussion of the possible results of bubble collisions suggests a spectrum ranging from what might be called ``Fatal" collisions to ``Perturbative" ones.  Fatal collisions would destroy all observers to their future, while Perturbative collisions would merely ``paint" their effect on the observation bubble.  Realistic collisions would fall in between these extremes.

Consider first a scenario in which Fatal (downward) bubbles can form
at rate $\lambda_{\rm fatal}$ and collide with our observation
bubble. Focusing on the $\tau=\tau_o$ spatial slice, on which we
presumably exist now, we must be at a position that has not yet
experienced such a collision. The unaffected volume fraction will be
$f_{\rm OK} = \exp[-\lambda_{\rm fatal}V_4(\xi_o)]$ (where $V_4$
measures the available past 4-volume for nucleations, which for
$\xi_o \gg 1$ is $V_4(\xi_o) \simeq 4\pi \xi_o H_{F}^{-4}$), and as
discussed in Sec.~\ref{sec-setup}, the 3-volume element goes as
$dV_3=4\pi H_{T}^{-3} \sinh^2(\xi_o)d\xi_o$.  Combining these, the
distribution in $\xi_o$, for $\xi_o \gg 1$, of volume unaffected by
fatal bubbles goes as
$$
dV_3 \ f_{\rm OK} \propto \exp[(2 - \frac{4\pi}{3} \lambda_{\rm fatal} H_{F}^{-4} )\xi_o] \ d\xi_o,
$$
For $\lambda_{\rm fatal} H_{F}^{-4} < \frac{3}{2 \pi}$ (which will be satisfied for any theory of eternal inflation) this diverges as $\xi_o \rightarrow \infty$, so we would expect {\em even the surviving regions} to be dominated by the largest $\xi_o$.

 Now, {\em if we assume} ourselves to be in a typical surviving region, there are two cases of interest. If we are in a bubble with $H_T \alt \lambda_{\rm fatal}^{1/2} H_{F}^{-1}$, then as time increases, we will have an increasing risk of being hit by Fatal bubble (as discussed in Sec.~\ref{sec-boostedangles}), and would expect such a collision after a cosmological time of order $\tau_o \sim \lambda_{\rm fatal}^{-1/2} H_{F}$.
Even if $H_T=0$, for exponentially small nucleation rates this can easily be a reassuringly long time;~\footnote{This analysis agrees with that of GGV, who essentially assumed that collisions are all Fatal and then found that we are unlikely to hit by such a bubble soon.} conversely, we can use our survival to rule out scenarios that include Fatal bubbles with  $\lambda_{\rm fatal}^{-1/2} H_{F} \agt 10\,$Gyr. If, instead, $H_T \agt \lambda_{\rm fatal}^{1/2} H_F^{-1}$, then all of the collision events likely to ever affect us happened in the distant past, and we will safely inhabit our unaffected region of the observation bubble, oblivious to the fact that fatal collisions may have occurred elsewhere.

Let us consider collisions that are Compatible but not Fatal, so that we might exist in at least part of the collision's future.  If this part is relatively small, or excludes the region that we are likely to be in, we might treat these bubbles as Fatal, and simply assume that we are not in the future of any of them.  If, on the other hand, we might exist in essentially all of the collision's future, we might treat them as Perturbative. If a theory predicts that at least one collision type is effectively Perturbative, then we can simply assume ourselves to be in a region unaffected by non-Perturbative bubbles, but should still expect to see Perturbative collisions to our past, following our derived distribution function. Determining whether a Compatible collision is effectively Fatal or Perturbative will be difficult, as it requires a detailed understanding of the collision's aftermath, and may also involve 'measure' issues to determine whether or not the (putative) observers in question are likely be in the perturbed or the destroyed part of the collision result.  (One cause for concern in this regard is that the $\xi_o \rightarrow \infty$ observers likely to see many collisions are very highly ``boosted".  Therefore even if an incoming bubble is almost perfectly Perturbative, this perturbation might be extremely dangerous to such a highly-boosted worldline. Another way to see this is to note that most collisions observed at early times by the ``boosted" observer in Fig.~\ref{fig-angles} come from very early cosmological times.)

In our analysis, we have concentrated on determining the region of the observer's sky that is {\em in principle} affected by (a set of) collision events. Further, we have used the bubble wall as the surface upon which the observer is examining the effects of collisions. This has allowed us to avoid making any assumptions about how collision products may travel inside of the observation bubble. However, the most relevant calculation is to determine the effects of bubble collisions on the post-tunneling equal-field surface, then in turn the observable effect on the last-scattering surface (and therefore in the CMB).  This will necessarily involve a better understanding of the physics involved in bubble collisions, an investigation that we reserve for future work.

That being said, we might speculate that the gross features of the distribution function on the last scattering surface will be similar to the analysis we have carried out, suggesting that bubble collisions would produce anisotropies and features on large angular scales in the CMB. Because of the bimodality of the distribution function, the subdominant peak around $\psi \simeq 0$ might also produce observable effects akin to point sources, but only if $\lambda \agt (H_TH_{F})^{-2}$ for some bubble type. These speculations must be put on much firmer ground before any conclusions can be drawn from current or future data.

\section{Discussion}
\label{sec-discuss}
In Sec.~\ref{sec-intro}, we outlined three conditions that must be met for there to be observable effects of bubble collisions in false-vacuum eternal inflation: Compatibility, Probability, and Observability. What do our results imply about these?

We have not gone beyond the general arguments concerning Compatibility given in Sec.~\ref{sec-intro}, except to note that incoming bubbles of higher vacuum energy are likely to be separated from us by a domain wall that accelerates away from us, greatly enhancing the likelihood that they will merely perturb the ``observation bubble." We have {\em not}, however, actually shown that bubbles with the requisite level of Compatibility are expected; it will be necessary to extend previous bubble-collision analyses~\cite{Hawking:1982ga,Bousso:2006ge,Blanco-Pillado:2003hq,Freivogel:2007fx} to answer this question decisively, as well as to assess the result of multiple bubble collisions affecting a single point inside the observation bubble.

Our main result is a calculation of the statistical distribution of
collisions coming from a direction $(\theta_n, \phi_n)$ that can
affect an angular scale $\psi$ on the 2-sphere defined by the
portion of the bubble wall causally accessible to an observer at
some instant in time, {\em assuming} that the incoming bubbles
merely perturb the observation bubble. The properties of this
distribution function depend upon the location of the observer
inside of the observation bubble, which we have evaluated in
complete generality, but there are two limiting cases of interest.

First, if we sit very far from the finite ``unaffected" region near the center of the bubble (defined by $\xi_o \alt \lambda H_{F}^{-4}$ in terms of the false-vacuum Hubble parameter $H_F$), then our results show that most collisions come from the direction of the bubble wall, happen at early observation times, and have a {\em large} angular scale $\psi \simeq 2 \pi$. If such bubble collisions are compatible with our observations, there is no reason to expect that they are not causally accessible to us.

Second, for an observer at any $\xi_o$, bubbles can potentially be encountered (or come into view) at late times $\tau_o \sim \lambda^{-1/2} H_{F}$  {\em if} $H_T \alt \lambda^{1/2} H_F^{-1}$. (Note that such values of $\lambda$ are large compared to typical exponentially suppressed nucleation rates, but still small compared to values that would allow percolation and thus preclude eternal inflation.)

Now consider Observability. One might have guessed that even if an
infinite number of bubbles collide with ours, they might be of
infinitesimal angular size on the sky, perhaps even taking up small
total sky fraction. Indeed this appears to be true for the small
scale, late-time collisions, but is {\em not} the case for the
early-time collisions -- which take up large angular scales.
Therefore, these early-time collisions will at least partially
satisfy the Observability criterion.

Assessing the other half of Observability (that the effects of the
collisions must survive inflation within the bubble) would, in the
context of eternal inflation, require both an accurate model of the
inflaton potential, and also a measure over transitions within this
potential so as to give a probability distribution over
e-foldings~\cite{Aguirre:2006na}. Neither is in hand but the present
results increase the importance of making progress in this area.

In addition, one must understand exactly how the effects of
early-time collisions would be imprinted on an observable like the
CMB. The surface of constant density corresponding to the beginning
of inflation will presumably be perturbed by collision products that
propagate into the bubble, and these effects translate into density
fluctuations on the surface of last scattering. Because we have only
treated the effects of collisions on the bubble wall, our analysis
is only a preliminary step to answering such detailed questions.
Nonetheless, it seems likely that  some basic features of the
distribution function, such as anisotropy and effects on large
angular scales, will persist.

In some sense, bubble collisions are the most generic prediction
made by false vacuum eternal inflation, independent of the
properties of the fundamental theory that may drive it. While
connecting this prediction to real observational signatures will
entail both difficult and comprehensive future work (and probably no
small measure of good luck), it appears worth pursuing. For a
confirmed observational signature of other universes, while
currently speculative even in principle, and probably far-off in
practice, would surely constitute an epochal discovery.

\begin{acknowledgments}
The authors wish to thank T. Banks, R. Bousso, B. Freivogel, and A.
Vilenkin for helpful discussions. MJ thanks the ARCS Foundation and
the Hierarchical Systems Research Foundation for support. MJ and AA
were partially supported by a ``Foundational Questions in Physics
and Cosmology" grant from the Templeton Foundation during the course
of this work.
\end{acknowledgments}

\appendix
\section{Triple intersection in the unboosted frame}\label{app-unboosted}\

In this appendix we solve directly for the coordinate angles denoting the boundaries of a collision on the Poincar\'e disk. We specialize to the case $H_T = H_F = H$, where it is possible to foliate the bubble interior with the flat slicing. Working in a plane of constant-$\phi$,\footnote{As before, we work with the convention where $-\pi < \theta < \pi$ to  cover full circles.} we are attempting to find the triple-intersection between three circles representing the observation bubble, the colliding bubble, and the past light cone of the observer, whose radii are given by
\begin{eqnarray}
\label{eq-flatrobs}
r_{obs} & = &  1 - e^{-H t}, \\
\label{eq-flatrcoll}
r_{coll} &=& e^{-H t_n} - e^{-H t}, \\
\label{eq-flatrplc}
r_{plc} &=&  e^{-H t} - e^{-H t_o}.
\end{eqnarray}
Using up the remaining symmetry of the problem we can assume that the observer is at $\theta_o=0$. The free parameters that must be specified are then the position at which the colliding bubble is nucleated $(t_n,r_n,\theta_n)$ and the position of the observer $(t_o, r_o)$ in terms of the flat slicing coordinates. The transformation between the open and flat slicing location of the observer is given by
\begin{equation}\label{trflop}
\begin{split}
r_o&=\frac{H^{-1} \sinh\xi_o\sinh\tau_o}{\cosh\tau_o+\cosh\xi_o\sinh\tau_o}\\
t_o&= H^{-1} \log(\cosh\tau_o+\cosh\xi_o\sinh\tau_o).
\end{split}
\end{equation}
The observation bubble introduces no new free parameters, since it is centered around the origin, and nucleates at $t=0$.

 We find it useful to parameterize time with $x\equiv 1-e^{-H t}$ (this way $r=x$ is the
observation bubble). It is straightforward to conclude that the three light-cones are
the set of points $(r(x,\theta),x,\theta)$ parameterized as follows:
\begin{itemize}
    \item {Observation Bubble future lightcone:
    \begin{equation}\label{ob}
    (r=x,\ x,\ \theta)\qquad 0\leq x\leq 1,\ -\pi\leq\theta\leq\pi
\end{equation}}
    \item {Observer's past lightcone:
    \begin{equation}\label{oplc}
    \begin{split}
    &(r_o\cos\theta\pm\sqrt{(x-x_o)^2-r_o^2\sin^2\theta},\ x,\
    \theta)\\
    &x\leq x_o,\ |\theta|\leq |\arcsin(\frac{x-x_o}{r_o})|
\end{split}\end{equation}}
    \item {New bubble future lightcone:
\begin{equation}\label{nbflc}
    \begin{split}
    &(r_n\cos(\theta-\theta_n)\pm\sqrt{(x_n-x)^2-r_n^2\sin^2(\theta-\theta_n)},\ x,\
    \theta)\\
    &x_n\leq x,\ |\theta-\theta_n|\leq
    |\arcsin(\frac{x_n-x}{r_n})|
\end{split}\end{equation}}
\end{itemize}
The triple intersection is the set of points belonging to all
three groups. Demanding first that
$1-x=r_o\cos\theta\pm\sqrt{(x-x_o)^2-r_o^2\sin^2\theta}$ and
repeating for
$1-x=r_n\cos(\theta-\theta_n)\pm\sqrt{(x_n-x)^2-r_n^2\sin^2(\theta-\theta_n)}$, then
solving for $x(\theta)$ we obtain
\begin{equation}\label{timesolbf}
2 x=\frac{r_o^2-x_o^2}{r_o \cos\theta-x_o}=\frac{r_n^2-x_n^2}{r_n
\cos(\theta-\theta_n)-x_n},
\end{equation} giving an equation for $\theta$:
\begin{equation}\label{bfeq}
\begin{split}
A&\cos\theta+B\sin\theta+C=0,\qquad{\rm where}\\
A&=r_o\bigl(x_n^2-r_n^2\bigr)-\cos\theta_n r_n
\bigl(x_o^2-r_o^2\bigr)\\
B&=-\sin\theta_n r_n
\bigl(x_o^2-r_o^2\bigr)\\
C&=x_n\bigl(x_o^2-r_o^2\bigr)-x_o\bigl(x_n^2-r_n^2\bigr).\\
\end{split}
\end{equation}
There are two solutions\footnote{The denominator $A^2+B^2$ never
vanishes because the observer and the nucleated bubble never sit
on the observation bubble wall. Also, notice that the symmetry in
$\phi$ is reflected in the fact that the positive solution for a
given $\theta_n$ is the negative solution for $-\theta_n$.} to Eq.
\ref{bfeq},
\begin{equation}\label{solbfeq}
\cos\theta_{1,2}=-\frac{\Bigl(AC\pm
B\sqrt{A^2+B^2-C^2}\Bigr)}{A^2+B^2}.
\end{equation}
One can now solve for the time of the intersection
by plugging $\theta_{1,2}$ into eq. \ref{timesolbf}. This gives
the coordinates of the two desired intersection events in the flat
slicing where the angle is measured from the origin. By spherical symmetry, these angles are the same as the coordinate angles  measured from the origin of the of the bubble interior as described by the open slicing coordinates. We can then use the angles $\theta_{1,2}$ to define the angle as
measured by the observer sitting at some open slicing coordinates
$(\xi_o,\tau_o,\theta_o=0)$ via Eq.~\ref{eq-unboostcospsi}.

\section{Effects of boosts on the bubble}
\label{app-boost}

In Sec.~\ref{sec-boostedview}, we used the symmetries of the one-bubble spacetime to justify performing a boost that would bring us to a frame where the observer is at the origin. Here, we explore the effects of this boost on the interior spacetime in greater detail.

In terms of the embedding coordinates, the transformation is given
by Eq.~\ref{eq-embeddingboost}. The first important property to note
is that the $X_4$ coordinate is invariant. In the open slicing,
surfaces of constant $X_4$ are surfaces of constant $\tau$, and so
we see that the boost preserves the open slicing time. The second
important property is that the observer at $(\xi_o, \tau_o,
\theta_o=0)$ is translated to the origin $(\xi_o'=0, \tau_o'=\tau_o,
\theta_o'=0)$ of the the boosted frame. From the relation for $X_0'$
in Eq.~\ref{eq-embeddingboost},
\begin{equation}
\cosh \xi_o' = \cosh \xi_o \left( \cosh \xi_o - \tanh \xi_o \sinh \xi_o  \right) = 1,
\end{equation}
and therefore $\xi_o' = 0$.

In Sec.~\ref{sec-computations}, we derived a formula for the observed angular scale of a collision event in both the boosted and unboosted frames. We now establish the invariance of this quantity by directly applying the transformation to Eq.~\ref{eq-unboostcospsi}. The angle $\theta$ in this equation corresponds to the angular position of the intersection on the null wall of the observation bubble (as defined by the origin in the unboosted frame), so using $\eta = T$, the boosted angle from Eq.~\ref{eq-closedboost} is:
\begin{equation}
\tan \theta' = \frac{\sin \theta}{\gamma \left( \cos \theta - \beta \right)}.
\end{equation}
In this frame, $\theta'$ can be identified as $\alpha$, the actual observed angle at which the boundary of the collision lies (which is used to find the total angular scale of the collision in Eq.~\ref{eq-unboostcospsi}). Solving for $\cos \theta'$,
\begin{equation}\label{eq-walltheta}
\cos \theta' = \frac{\sinh \xi_o - \cos \theta \cosh \xi_o}{ \sqrt{\sin^2 \theta + \left( \sinh \xi_o - \cos \theta \cosh \xi_o \right)^2  }},
\end{equation}
and expanding into exponentials reveals that this expression is in fact equal to Eq.~\ref{eq-unboostcospsi}, as evidenced by:
\begin{eqnarray}
\cos \alpha &=& \cos \theta'  \\
&& \nonumber \\
&=& -\frac{1+2e^{i\theta} + e^{2i\theta} + e^{2i \xi_o} - 2 e^{i\theta + 2 \xi_o} + e^{2i\theta+2 \xi_o} }{1+2e^{i\theta} + e^{2i\theta} - e^{2i \xi_o} + 2 e^{i\theta + 2 \xi_o} - e^{2i\theta+2 \xi_o}} \nonumber
\end{eqnarray}
In the Poincar\'e disk representation, using the hyperbolic law of cosines, this implies that all of the angles in the triangle composed of (and therefore the lengths between) the observation point, the unboosted position of the origin, and the edge of the collision, remain invariant under the boost. More generally, the distance between any two points on the disc will be invariant under the boost (as one can check on a point-by-point basis), and so we can identify the boost as a pure translation in the open coordinates.

\bibliography{collisions}

\end{document}